\documentclass[english,aps,preprint,nofootinbib]{revtex4}
\usepackage{graphicx}

\makeatletter

\providecommand{\tabularnewline}{\\}
\newcommand{\lyxdot}{.}

\makeatletter

\newcommand{\bee}{\begin{equation}}
\newcommand{\ee}{\end{equation}}
\newcommand{\beea}{\begin{eqnarray}}
\newcommand{\eea}{\end{eqnarray}}

\preprint{HU-EP 08/15; SFB/CPP-08-27}

\makeatother

\usepackage{babel}
\makeatother

\begin{document}

\title{Reweighting towards the chiral limit}

\author{Anna Hasenfratz}
\email{anna@eotvos.colorado.edu}
\affiliation{Department of Physics, University of Colorado, Boulder, CO-80309-390}

\author{Roland Hoffmann}
\email{hoffmann@pizero.colorado.edu}
\affiliation{Bergische Universit\"at Wuppertal, Gaussstr.~20, 42219 Wuppertal,
Germany}

\author{Stefan Schaefer}
\email{sschaefer@physik.hu-berlin.de}
\affiliation{Institut f\"ur Physik, Humboldt Universit\"at, Newtonstr.~15, 12489
Berlin, Germany}

\begin{abstract}
We propose to perform fully dynamical simulations
at small quark masses by reweighting in the quark mass. This approach avoids
some of the technical difficulties associated with direct simulations
at very small quark masses. We calculate the weight factors stochastically,
using determinant breakup and low mode projection to reduce the statistical
fluctuations. We find that the weight factors fluctuate only moderately
on nHYP smeared dynamical Wilson-clover ensembles, and we could successfully
reweight $16^{4}$, ($1.85$fm)$^{4}$ volume configurations from
$m_{q}\approx20$MeV to $m_{q}\approx5$MeV quark masses, reaching
the $\epsilon-$regime. We illustrate the strength of the method by
calculating the low energy constant $F$ from the $\epsilon-$regime
pseudo-scalar correlator.
\end{abstract}
\maketitle

\section{Introduction}

The steady progress of simulation techniques over the last decade
(see e.g. \cite{Morningstar:2003gk,Luscher:2005rx,Urbach:2005ji,Takaishi:2005tz,Hasenfratz:2007rf})
as well as new insights into the reasons for algorithmic failures
\cite{DelDebbio:2005qa} have profoundly altered the status of lattice
QCD: With the latest generation of supercomputers essentially all
$p$-regime points, including the point of physical quark masses, have
become accessible to direct simulation. However, in the small quark
mass regime the challenges are still considerable:

\begin{itemize}
\item Large volumes are needed for the stability of the algorithms when
Wilson fermions are used \cite{DelDebbio:2005qa}.
\item Auto-correlation times increase dramatically towards the chiral limit
\cite{Aoki:2008tq}. 
\item Statistical fluctuations of fermionic correlators become difficult
to estimate since configurations with large contributions become rare
as small Dirac modes are more and more suppressed.
\end{itemize}
A possible solution to evade these problems is to avoid generating
an ensemble with the fermionic weight of the small target quark mass
but instead simulate a heavier quark  and reweight to the desired
ensemble.

This approach solves all of the above mentioned issues: The algorithm
is more efficient at the larger mass, and smaller volumes will be
sufficient from the algorithmic point of view. At a larger quark mass
the region of small Dirac eigenvalues is oversampled with respect
to the target distribution and thus observables that receive large
contributions there (e.g. pseudo-scalar correlators) will be better
estimated.
Previously, the Polynomial Hybrid Monte Carlo algorithm
\cite{Frezzotti:1997ym} has been used as an alternative way
to achieve such an oversampling \cite{DellaMorte:2004hs,Jansen:2007rx}.

Since the variance of an observable $O$ is again a field theoretical
observable, its statistical error on an importance sampled ensemble
depends only on the auto-correlation of $O$. On
the other hand when reweighting is employed the error is not given
by the quantum mechanical variance $\langle O^{2}w\rangle-\langle Ow\rangle^{2}$,
with $w$ being the normalized reweighting factor, but rather the
statistical variance $\langle O^{2}w^{2}\rangle-\langle Ow\rangle^{2}$
which depends on the variance of the reweighting factor itself as
well as its correlation with the observable of interest. If $O$ and
$w$ are (strongly) anti-correlated, this controls and limits the
statistical error on the reweighted ensemble.

Nevertheless, if the overlap (in configuration space) of the generated
and desired ensembles is insufficient, reweighting will break down
due to the fluctuations of the reweighting factor. This limits the
range of quark mass values that can be bridged by reweighting.

In this paper we work with two degenerate flavors of Wilson type fermions,
though generalization to other fermionic actions is straightforward.
If we have an ensemble of configurations $\{U_{i}\}$ generated at
bare mass $m_{1}$ with the Dirac operator $D_{1}=D(U;\, m_{1})$,
we can reweight it to the ensemble that corresponds to bare quark
mass $m_{2}$ by assigning to each configuration a weight factor
\begin{equation}
w_{i}\propto\det{\rm  }\frac{D_{2}^{\dagger}[U_{i}]D_{2}[U_{i}]}{D_{1}^{\dagger}[U_{i}]D_{1}[U_{i}]}\,,
\label{eq:w-1}
\end{equation}
and calculating expectation values as
\begin{equation}
\langle O\rangle_{2}=\frac{\sum_{i}w_{i}O[U_{i}]}{\sum_{i}w_{i}}\,.
\label{eq:unused1}
\end{equation}
Since the reweighting factors and their fluctuations will increase with the volume,
reweighting
becomes inefficient on very large lattices. In practice  we find that the
fluctuations can be controlled and quite large volumes can be reweighted.
Nevertheless, reweighting is a technique that is most useful at small
quark masses and moderate volumes -- like in $\epsilon$-regime calculations.

There are two sources of fluctuations for the weights. One is due
to the small eigenmodes of the Dirac operator. These physical infrared
eigenmodes contribute to the weight factor as
\begin{equation}
\log(w_{i})\Big\Vert_{{\rm low\, modes}}=(m_{2}\!-\! m_{1})\sum_{\lambda}
\frac{1}{\lambda+m_{1}}+{O}((m_{2}\!-\! m_{1})^{2})\,.
\label{eq:low_modes}
\end{equation}
 The suppression of configurations due to the small eigenvalues is
physical. The weight factors control exceptionally large contributions
to quark propagators that arise on configurations with small eigenmodes.
Reweighting in fact reduces the statistical fluctuations of many observables
when compared to the partial quenched case. The ultraviolet, large
eigenvalue modes are not physical, but due to their large number they
can dominate the fluctuations. Some of these fluctuations can be removed
by smoothing, and with nHYP smeared Dirac operators \cite{Hasenfratz:2007rf} reweighting
is possible also in bigger volumes. However, even on an nHYP smeared
gauge background, the UV fluctuations are still large. They are also
closely correlated with the fluctuations of the nHYP smeared plaquette,
giving us yet an other option to reduce the noise by absorbing it
into the gauge action: Including a term proportional to the smeared
plaquette in the Lagrangian reduces the UV fluctuations of the weight
factors. This latter reduction is not essential, especially at smaller
mass shifts, but extends the reach of the method at the expense of
introducing a (very) small pure gauge term to the action.

Calculating the determinant in Eq. (\ref{eq:w-1}) to any reasonable
accuracy can be very costly. Fortunately it is not necessary to do
that, a stochastic estimator is sufficient. In Sect. \ref{sec:Stochastic-reweighting}
we describe the stochastic reweighting, and several of its improvements.
Sect. \ref{sec:Numerical-tests} describes the numerical tests and
efficiency of the reweighting technique, and in Sect. \ref{sec:Physics-results}
we present physics results using reweighted configurations.

\section{Stochastic reweighting\label{sec:Stochastic-reweighting}}

When the Dirac operator corresponds to Wilson or clover fermions it
has the form%
\footnote{here $M$ is a suitable combination of the hopping and clover terms}
\begin{equation}
D[U]=1-\kappa M[U]\;,
\label{eq:unused2}
\end{equation}
where $\kappa$ is the hopping parameter $\kappa=(2m+8)^{-1}$. Reweighting
a configuration from $\kappa_{1}$ to $\kappa_{2}$ requires a weight
factor
\begin{eqnarray}
w & = & {\rm det}\frac{D_{2}^{\dagger}D_{2}}{D_{1}^{\dagger}D_{1}}={\rm det}^{-1}(\Omega)\,,\nonumber \\
\Omega & = & D_{2}^{-1}D_{1}D_{1}^{\dagger}(D_{2}^{\dagger})^{-1}\,.
\label{eq:Omega}
\end{eqnarray}
The determinant can be calculated as an expectation value
\[
w=\frac{\int\!\mathcal{D}\xi\, e^{-\xi^{\dagger}\Omega\xi}}{\int\!\mathcal{D}\xi\, e^{-\xi^{\dagger}\xi}}=\langle e^{-\xi^{\dagger}(\Omega-1)\xi}\rangle_{\xi}\;,
\]
but obtaining a reliable estimate for $w$ is expensive, especially
when $\Omega$ is not close to one.

An alternative way is to calculate only a stochastic estimator of
the true weight factor and do the average over the $\xi$ fields together
with the configuration average. A similar approach is used in the
stochastic global Monte Carlo update \cite{Hasenfratz:2002jn,Alexandru:2002jr,Hasenfratz:2005tt}.
We start by writing the expectation value of an operator $O[U]$ at
mass $m_{2}$ as
\begin{eqnarray}
\langle O\rangle_{2} & = & \frac{1}{Z_{2}}\int\!\mathcal{D}Ue^{-S_{g}}{\rm det}(D_{2}^{\dagger}D_{2})\, O[U]\nonumber \\
 & = & \frac{1}{Z_{2}}\int\!\mathcal{D}Ue^{-S_{g}}{\rm det}(D_{1}^{\dagger}D_{1}){\rm det}^{-1}(\Omega)\, O[U]\label{eq:unused3}\\
 & = & \frac{Z_{1}}{Z_{2}}\langle O[U]e^{-\xi^{\dagger}(\Omega-1)\xi}\rangle_{1,\xi}\,,\nonumber
\end{eqnarray}
where
\begin{eqnarray}
\frac{Z_{2}}{Z_{1}} & = & \frac{\int\!\mathcal{D}Ue^{-S_{g}}{\rm det}(D_{2}^{\dagger}D_{2})}{\int\!\mathcal{D}Ue^{-S_{g}}{\rm det}(D_{1}^{\dagger}D_{1})}\label{eq:unused4}\\
& = & \langle e^{-\xi^{\dagger}(\Omega-1)\xi}\rangle_{1,\xi}\nonumber
\end{eqnarray}
and the expectation value is with respect to both the $U$ and $\xi$
fields at $m_{1}$. If we consider the configuration ensemble $\{U_{i},\xi_{i}\}$,
with the gauge configurations from the original sequence and $\xi_{i}$
generated independently with weight $e^{-\xi^{\dagger}\xi}$, the
expectation value becomes
\begin{equation}
\langle O\rangle_{2}=\frac{\sum_{i}O[U_{i}]e^{-\xi_{i}^{\dagger}(\Omega[U_{i}]-1)\xi_{i}}}{\sum_{i}e^{-\xi_{i}^{\dagger}(\Omega[U_{i}]-1)\xi_{i}}}\,,
\label{eq:weighted_exp}\end{equation}
i.e. the weight factors are replaced by a single estimator
\begin{equation}
s_{i}=e^{-\xi_{i}^{\dagger}(\Omega[U_{i}]-1)\xi_{i}}\,.
\label{eq:weight}
\end{equation}
The obvious advantage of the stochastic approach is that the averages
of the noise sources and the gauge fields commute. 
Without introducing any systematic error, we therefore need
only one estimator of the weight on each configuration.
The disadvantage is that a single estimator might fluctuate too much
and introduce large statistical errors in Eq. (\ref{eq:weighted_exp}).
Fortunately there are several methods that can reduce the fluctuations
of $s_{i}$ to acceptable levels.

\subsection{Improving the stochastic estimator}

In this section we describe two methods, the determinant breakup
\cite{Hasenbusch:1998yb,Hasenbusch:2001ne}
and the low mode subtraction that we use in calculating the stochastic
weight factors. Both methods were used in \cite{Hasenfratz:2005tt}
in a different context. It will be useful for the discussion to rewrite
Eq. (\ref{eq:Omega}) as
\begin{equation}
\Omega=\big((1-x)(D_{2}^{\dagger})^{-1}+x\big)\big((1-x)D_{2}^{-1}+x\big)\,,
\label{eq:Omega2}
\end{equation}
where $x=\kappa_{1}/\kappa_{2}$. We consider reweighting to smaller
quark masses so $x<1$, though everything works for $x>1$ as well.
At leading order $(1-x)\propto(m_{1}-m_{2})$, i.e. apart from a gauge
field independent additive constant $x^{2}$, $\Omega\propto(m_{1}-m_{2})$.
The exponent of the stochastic estimator in Eq. (\ref{eq:weight})
now can be written as
\begin{equation}
\xi^{\dagger}(\Omega-1)\xi=(1-x)^{2}\xi^{\dagger}(D_{2}^{\dagger})^{-1}D_{2}^{-1}\xi+x(1-x)\xi^{\dagger}
\big((D_{2}^{\dagger})^{-1}+D_{2}^{-1}\big)\xi+(x^{2}-1)\xi^{\dagger}\xi\,,
\label{eq:stoch_estimator}
\end{equation}
which requires one inversion of the Dirac operator $D_{2}$ on $\xi$.

\subsubsection{Determinant Breakup}

The determinant breakup is based on Ref. \cite{Hasenbusch:2001ne}
and has been used extensively in dynamical simulations. The idea is
to break up the interval $(\kappa_{1}\to\kappa_{2})$ to $N$ sections
$\kappa_{1}\to\kappa_{1}+\Delta\kappa\to....\to\kappa_{1}+N\Delta\kappa=\kappa_{2}$,
and write the determinant as the product of $N$ terms, each on a
$\Delta\kappa$ interval. Thus the stochastic estimator takes the
form
\begin{equation}
\exp\Big(-\sum_{n=1}^{N}\xi_{n}^{\dagger}(\Omega_{n}-1)\xi_{n}\Big)\;,
\label{eq:det_breakup}
\end{equation}
where $\Omega_{n}$ is the analogue of Eq. (\ref{eq:Omega}) on the
$\kappa_{n}\to\kappa_{n+1}$ interval. Since the operators $\Omega_{n}$
are much closer to one then the original $\Omega$, the estimator
in Eq. (\ref{eq:det_breakup}) has reduced fluctuations. Calculating
it on $N$ intervals requires $N$ applications of $D^{-1}$, so it
is expensive. On the other hand it predicts a stochastic weight factor
for the configuration at any intermediate $\kappa_{n}$ value, so
the same calculation can be used to reweight to many different mass
values.

\subsubsection{Separating the low eigenmodes}

Eqs.~(\ref{eq:low_modes}) and (\ref{eq:stoch_estimator}) clearly
show that the low eigenmodes of the Dirac operator not only give large
contributions to the reweighting factor but they can dominate the
stochastic fluctuations as well. This latter problem can be reduced
by explicitly calculating the low eigenmodes and removing them from
the stochastic estimator. It is not necessary to work with the exact
eigenmodes of $\Omega$, subtracting approximate eigenmodes works
as well.

Assume that $P$ is an arbitrary but Hermitian projection operator,
$P^{2}=P$, $P^{\dagger}=P$, and $\bar{P}=1-P$ is the complementary
projector. Any operator can be decomposed as
\begin{eqnarray*}
\Omega & = & \left(\begin{array}{cc}
P\Omega P & P\Omega\bar{P}\\
\bar{P}\Omega P & \bar{P}\Omega\bar{P}\end{array}\right)\\
& = & \left(\begin{array}{cc}
1 & 0\\
T & 1\end{array}\right)\left(\begin{array}{cc}
P\Omega P & 0\\
0 & Q\end{array}\right)\left(\begin{array}{cc}
1 & R\\
0 & 1\end{array}\right)
\end{eqnarray*}
with $T=\bar{P}\Omega P(P\Omega P)^{-1}$, $R=(P\Omega P)^{-1}P\Omega\bar{P}$ and
\begin{equation}
Q=\bar{P}\Omega\bar{P}-\bar{P}\Omega P(P\Omega P)^{-1}P\Omega\bar{P}\,.
\label{eq:proj2}
\end{equation}
Now the determinant can be written as
\begin{equation}
{\rm det\,}\Omega={\rm det\,}(P\Omega P)\,{\rm det\,}Q\,,
\label{eq:projected_det}
\end{equation}
so the sectors projected by $P$ and $\bar{P}$ are  separated.
If the projection operator $P$ is built form the eigenvectors of
$\Omega$, the second term in Eq. (\ref{eq:proj2}) vanishes and
$Q=\bar{P}\Omega\bar{P}$.
If $P$ is built only from approximate eigenmodes, both terms in
$Q$ are present but the second term gives only a small correction.
We will refer to it in the following as correction term.

\emph{Separating Dirac eigenmodes:} Due to the $\gamma_{5}$ Hermiticity
of the Wilson Dirac operator the eigenvalues of $D$ come in complex
conjugate pairs and the eigenvectors are $\gamma_{5}$ orthogonal.
The eigenvectors of the massless operator
$D_{0}|v_{\lambda}\rangle=\lambda|v_{\lambda}\rangle$
are the eigenvectors of the massive one as well, with eigenvalues
$\lambda+m$. The $\gamma_{5}$ orthogonality implies that one can
normalize the eigenvectors such that
$\langle v_{\lambda'}|\gamma_{5}|\,v_{\lambda^{*}}\rangle=\delta_{\lambda'\lambda}$
and the operator
\[
P_{\lambda}=|v_{\lambda}\rangle\langle v_{\lambda^{*}}\!|\,\gamma_{5}
\]
is a projector. Since $P_{\lambda}$ is not Hermitian, the discussion
of the previous paragraph does not apply. Nevertheless if $P_{\lambda}$
is built form the eigenvectors of Dirac operator, Eq. (\ref{eq:projected_det})
is valid with vanishing correction term. If $P$ projects to a few
low energy modes of $D_{2}$, the first term, $ $${\rm det}(P\Omega P)$,
can be calculated explicitly, and the second one, ${\rm det}\bar{(P}\Omega\bar{P})$,
can be estimated stochastically using Eq.~(\ref{eq:stoch_estimator}).
Since the eigenvectors of the massless Dirac operator can be used
at all mass values, the overhead of the low mode subtraction is a
one time calculation of the eigenmodes. If those modes are subtracted
during the inversion, some of the additional cost can be recouped
by the improved convergence as well.

\emph{Separating Hermitian eigenmodes}: An alternative approach is
to construct the projection operator from the eigenmodes of the Hermitian
operator \[
H=\gamma_{5}D\;.\]
Eqs. (\ref{eq:Omega}) and (\ref{eq:Omega2}) in terms of the Hermitian
operator become
\begin{eqnarray}
\Omega & = & H_{2}^{-1}H_{1}^{2}H_{2}^{-1}\nonumber \\
& = & (1-x)^{2}H_{2}^{-2}+x(1-x)(\gamma_{5}H_{2}^{-1}+H_{2}^{-1}\gamma_{5})+x^{2}\,.
\label{eq:Omega_herm}
\end{eqnarray}
Now the projector is Hermitian, but it is not constructed from the
eigenmodes of $\Omega$ so the correction term in Eq. (\ref{eq:proj2})
is necessary.

Working with $n$ normalized eigenvectors of $H_{2}|w_{i}\rangle=\eta_{i}|w_{i}\rangle$
the projector is $P=\sum_{i=1}^{n}|w_{i}\rangle\langle w_{i}|$ ,
and \[
P\Omega P=\omega_{ij}|w_{i}\rangle\langle w_{j}|\]
is an $n\times n$ matrix with coefficients
\begin{equation}
\omega_{ij}=(1-x)^{2}\frac{\delta_{ij}}{\eta_{i}^{2}}+
x(1-x)\langle w_{i}|\gamma_{5}|w_{j}\rangle({\eta_{i}^{-1}}+
{\eta_{j}^{-1}})+x^{2}\delta_{ij}\;.
\label{eq:11}
\end{equation}
Both the determinant and the inverse of $P\Omega P$ is easily calculable.
The leading term of the stochastic estimator, $\bar{P}\Omega\bar{P}$
requires the evaluation of $H_{2}^{-1}\bar{P}|\xi\rangle$. Since
the inversion is done on the subspace that is orthogonal to the low
eigenmodes, this can be considerably faster than calculating
$H_{2}^{-1}|\xi\rangle$.
The correction term does not require a new inversion if one uses the
identity $H_{2}^{-1}|w_{i}\rangle=1/\eta_{i}|w_{i}\rangle$.

The eigenmodes of $H_{2}$ have to be recalculated for every $\kappa$
interval. While the initial calculation of the eigenmodes is usually
expensive, changing $\kappa$ by a small amount does not effect the
eigenmodes much, and starting from nearly correct eigenmodes the calculation
converges fast. An alternative is to use the eigenmodes of $H$ of
the first interval throughout, but than one needs to calculate
$H_{2}^{-1}|w_{i}\rangle$
on every interval. The inversion on the low modes is expensive and
we found recalculating the eigenmodes on each interval is a better
option.

\section{Numerical tests\label{sec:Numerical-tests}}

\begin{figure}
\includegraphics[scale=0.4]{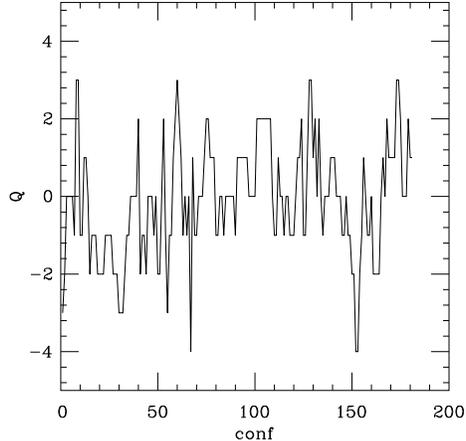}

\caption{The topological charge of the original configurations.
\label{fig:The-topological-charge}}

\end{figure}

We have tested the reweighting on a set of $16^{4}$ configurations
generated with a 2-flavor nHYP clover action. At the original parameter
values, $\beta=7.2$, $\kappa_{1}=0.1278$, the lattice spacing is
$a=0.115(3)\,$fm  and the PCAC quark mass is
$m_{PCAC}\approx20\,$MeV. We have 180 thermalized configurations separated
by 5 trajectories. The topological charge, as measured by an overlap
operator based on nHYP Wilson fermions ($R_{0}=0.6$ and no clover
term \cite{Hasenfratz:2007iv}) fluctuates evenly, suggesting that
the auto-correlation of these lattices is small even for this traditionally
slowly changing quantity (Figure \ref{fig:The-topological-charge}).

\begin{figure}
\includegraphics[scale=0.4]{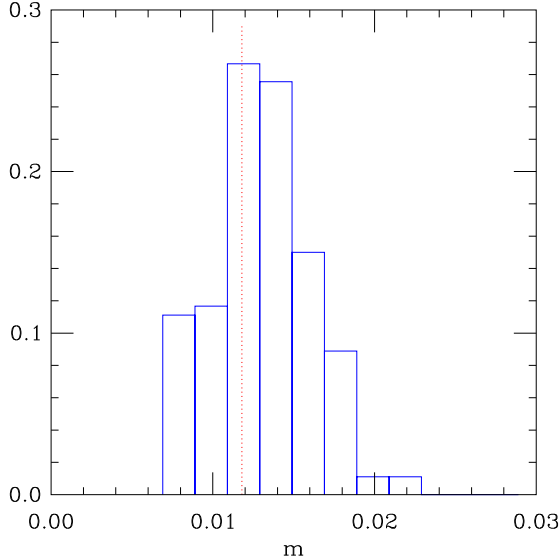}

\caption{The gap of the Hermitian Dirac operator of the original $16^{4}$,
$\kappa=0.1278$, $\beta=7.2$ configurations. The dashed line indicates
the PCAC quark mass.\label{fig:gap-of-H}}

\end{figure}

Figure \ref{fig:gap-of-H} shows the Hermitian gap, i.e. the histogram
of the absolute value of the lowest Hermitian eigenmode of the
configurations~\cite{DelDebbio:2005qa}.
Direct simulations are safe when the left edge of the gap is far from
zero. In our case one could probably lower the quark mass to about
$am_{q}=0.009$, but not much more. With reweighting, on the other
hand, one can go to much lower quark masses. Configurations with near
zero eigenvalues will be suppressed, just as they should be, so exceptional
configurations do not cause problems. Of course at some point one
has to worry about the chiral symmetry breaking lattice artifacts
of such actions. %

We reweighted the ensemble up to $\kappa_{2}=0.1281$, or $m_{PCAC}\approx5\,$MeV.
At this value there is one configuration in the ensemble with negative,
and a few with very small real (Dirac) eigenvalue. Reweighting to
such a small mass is interesting not necessarily for its physical
importance but rather to see the suppression of the determinant at
work. Since the stochastic reweighting automatically gives the weights
at intermediate mass values, we have done weighted spectrum calculations
at $\kappa=0.1279$ and $\kappa=0.1280$ as well. These values do
not cut into the gap and give physically more reliable results.

Our goal with these configurations is to probe the $\epsilon$-regime
with Wilson fermions. Already the configurations
at $\kappa=0.1278$ are controlled by the finite volume. From the
PCAC quark mass and the pseudo-scalar correlator on these runs, and
from our previous $16^{3}\times32$ dynamical runs \cite{Schaefer:2007dc,Hoffmann:2007nm}
we estimate $m_{\pi}L\approx2.8(1)$ on these $16^{4}$ lattices.
In the $\epsilon-$regime the eigenmodes of both the Dirac and Hermitian
Dirac operators are pushed away from zero, while the finite volume
has little effect on the PCAC quark mass. The ratio of the median
of the gap, $\bar{\mu}$, and $m_{PCAC}$ increases as one approaches
the $\epsilon-$regime. In infinite volume the ratio is the renormalization
factor $Z_{A}$, which we expect to be near 1 with nHYP fermions.
On large but finite volume the ratio might not have any physical meaning,
but there is indication that in practice it remains close to $Z_{A}$
\cite{DelDebbio:2005qa}. Our large volume runs give $\bar{\mu}/m_{PCAC}\approx0.82$,
while the ratio here is $1.1$, showing the effect of the finite volume.
(In \cite{Hasenfratz:2007rf} we quoted $\bar{\mu}/m_{PCAC}\approx0.91$.
Those runs on $12^{3}\times24$ lattices with $La=1.5$fm also have
strong finite volume effects even with the relatively heavy $m_{q}=70\,$MeV
quark mass.)

\subsection{Reweighting with Hermitian eigenmodes}

With Hermitian subtraction one has to balance the cost of calculating
the low energy eigenmodes with the improved convergence of the conjugate
gradient iteration. The cost of the latter is proportional to the
number of intervals between the starting and ending $\kappa$ values,
which in turn determines the statistical fluctuations of the stochastic
estimator due to the ultraviolet modes.

We have found that removing more than 6 eigenmodes did not substantially
increase the convergence of the inversion, nor did it decrease the
fluctuations of the estimator. Subtracting only 3 eigenmodes resulted
in a significantly more expensive inversion that quickly overtook the
cost of calculating the eigenmodes. Therefore we have settled on subtracting
6 Hermitian eigenmodes. Next we considered the optimal number of steps in
the determinant
breakup. This should be such that the stochastic fluctuations of the
weight factors are small compared to the fluctuations between the
weight factors of the different configurations. We found that 99 intervals
between $\kappa=0.1278$ and $\kappa=0.1281$ was more than sufficient
to achieve that. Going to smaller breakup might have worked but at
some point the start-up cost of the eigenmode calculations dominate
the cost.

For the computation of the eigenvalues and eigenvectors of the Hermitian
Dirac operator we use the Primme package of McCombs and Stathopoulos
\cite{primme1,primme2}.

\subsection{Reweighting with Dirac eigenmodes}

We have also tested reweighting with Dirac eigenmodes. Using the ARPACK
package we calculated 20 eigenmodes and separated $\approx\,$16 real
or complex conjugate pairs. While these eigenmodes work on every $\kappa$
interval, the conjugate gradient inversion is still expensive (we
project on $D_{2}$ modes but invert the operator $D_{2}^{\dagger}D_{2}$)
and we found this approach more expensive than removing Hermitian
eigenmodes. Our tests were done with single precision Dirac eigenmodes
and it is quite possible that with better eigenmodes the Dirac
eigenmode separation becomes competitive.

\subsection{Reweighting factors}

In this section we present results obtained using Hermitian eigenmode
separation. %
\begin{figure}
\includegraphics[scale=0.6]{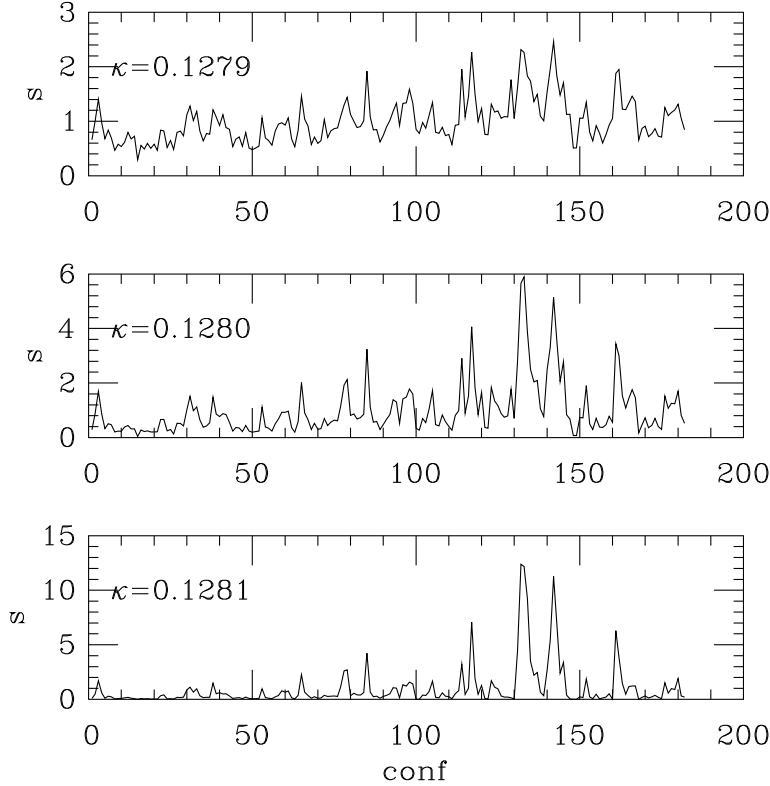}

\caption{Reweighting factors at $\kappa=0.1279$, $0.1280$ and $0.1281$ on
180 $16^{4}$ configurations. The reweighting factors are normalized
such that their average is $\langle s\rangle=1$.
\label{fig:Reweighting-factors}}

\end{figure}

Figure \ref{fig:Reweighting-factors} shows the reweighting factors
at three different $\kappa$ values starting form the original $\kappa=0.1278$.
As expected, they fluctuate more and more as we reweight to smaller
and smaller masses, and the last case, $\kappa=0.1281$, is just
barely acceptable. At that point there are several configurations
that have very small Hermitian eigenvalues, and the extreme suppression
of those configurations is evident. 

\begin{figure}
\includegraphics[scale=0.6]{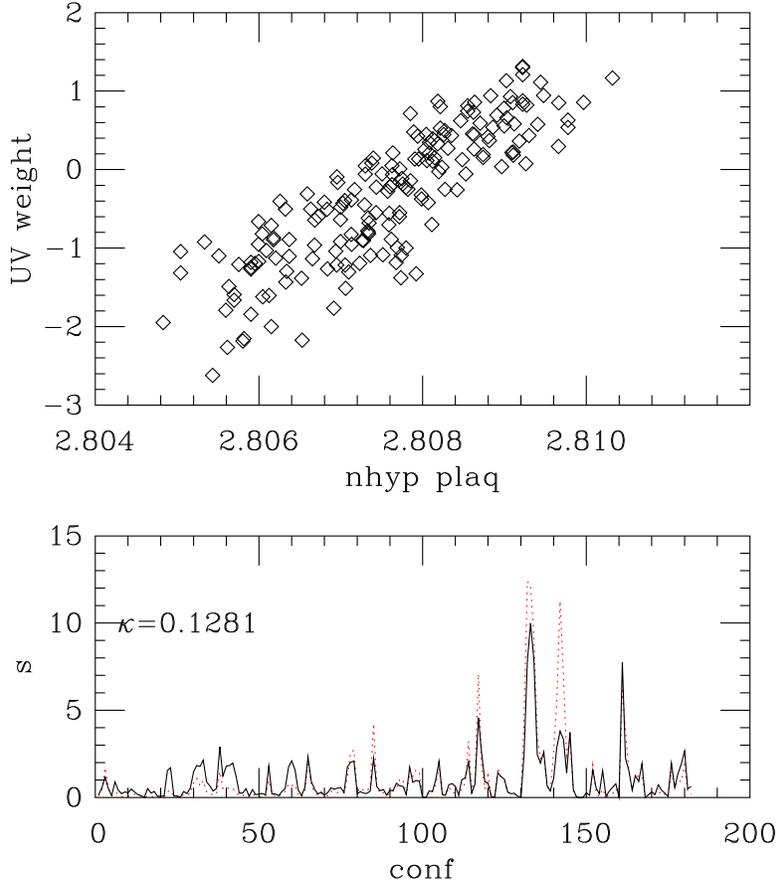} 

\caption{The logarithm of the UV part of the reweighting factor versus the
smeared nHYP plaquette (top panel), and the reweighting factors after
the observed correlation is removed by an nHYP plaquette term in the
action (lower panel). The dashed red line in the lower panel is the
original reweighting factor. The data is for $\kappa=0.1281$.
\label{fig:nHYP-correction}}

\end{figure}

There is another source for the large fluctuations, the ultraviolet
noise. The fluctuation of the nHYP plaquette is a good representative
of the UV noise, and it correlates closely with the UV part of the
weight factor (top panel of Figure \ref{fig:nHYP-correction} ). We
define the UV part as the weight factor without the explicitly calculated
low eigenmodes, i.e. the value determined by the stochastic process.
While this definition is not unique, it captures the correlation with
the nHYP plaquette and suggests that at least some of the UV noise
can be removed by introducing an nHYP plaquette term in the new action
by reweighting from $S_{1}=S_{g}-\ln\det(D_{1}^{\dagger}D_{1})$ to
\begin{equation}
S_{2}=S_{g}-\ln\det(D_{2}^{\dagger}D_{2})+\frac{\tilde{\beta}}{3}
\sum_{p}(3-{\rm Re\, Tr\,}U_{p,{\rm nHYP}})\,.\label{eq:nhyp_reweight}\end{equation}
 The correlation shown in Figure \ref{fig:nHYP-correction} can be
captured by an nHYP plaquette term with $\tilde{\beta}=-0.00133$ coefficient
at $\kappa=0.1281$, or $\tilde{\beta}=-0.00116$ at $\kappa=0.1280$.
The new term is a local, pure gauge term with very small coefficient.
While such a term is not necessary, it does help in reducing the fluctuations.
The lower panel of Figure \ref{fig:nHYP-correction} shows the reweighting
factors at $\kappa=0.1281$ both with (solid line) and without (dotted
line) an nHYP plaquette term in the action.

Of course the variance of the reweighting factor is only one aspect.
The real test is how they combine with observables to give fully unquenched
results. We present two examples here, both for reweighting to $\kappa=0.1280$.
\begin{figure}
\includegraphics[clip,scale=0.8]{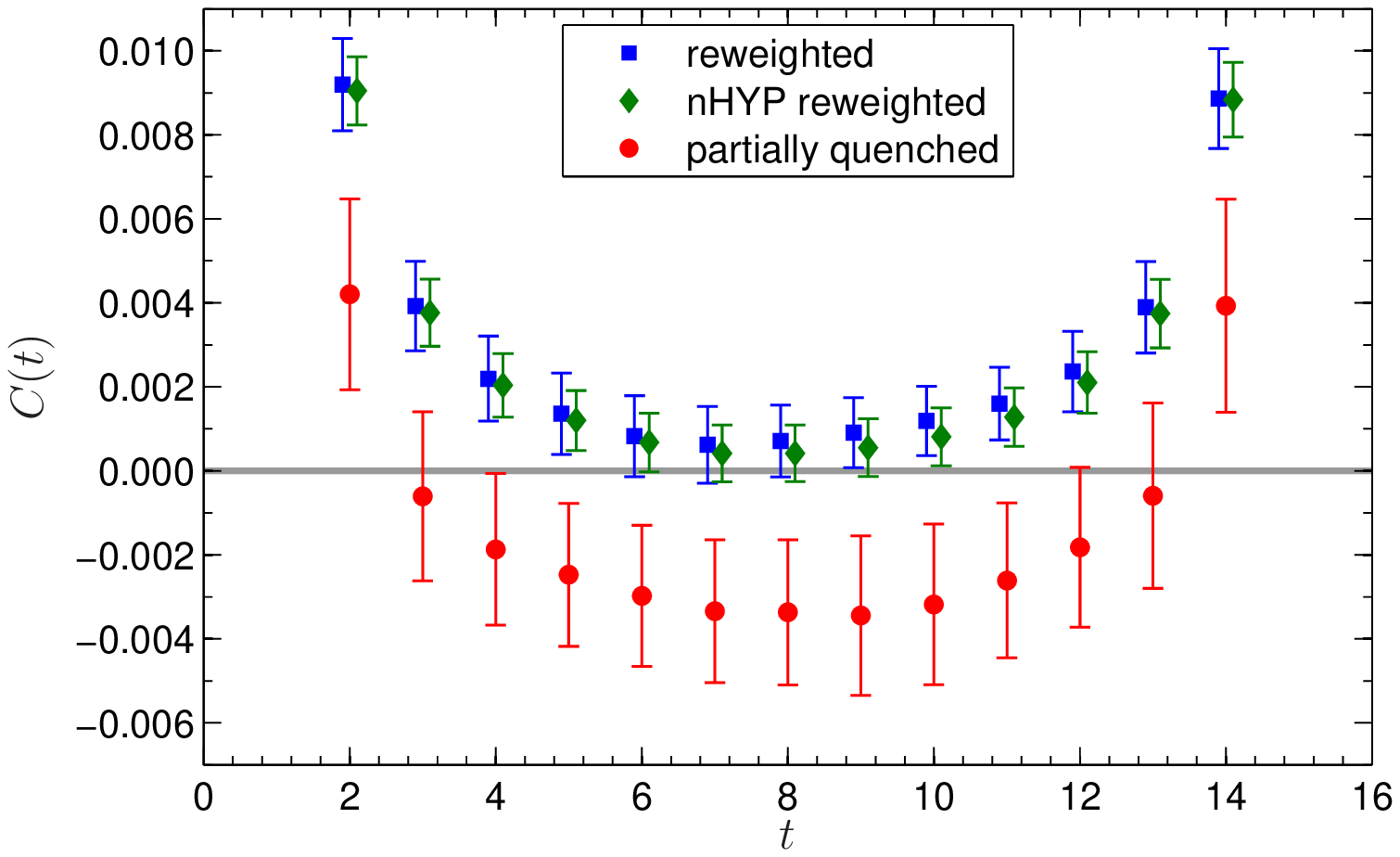} 

\caption{The scalar correlator at $\kappa=0.1280$ for the partially quenched,
reweighted and nHYP reweighted ensembles.\label{fig:The-scalar-correlator}
The reweighted data points are slightly offset for clarity.}

\end{figure}

Figure \ref{fig:The-scalar-correlator} shows the scalar correlator
for the partially quenched, reweighted and nHYP reweighted ensembles.
The partially quenched data shows that the correlator becomes negative,
a well known {}``artifact'' of partial quenching. This disease is
cured by reweighting. On the fully dynamical ensembles the propagator
is positive with both kinds of reweighting. In fact the two reweighted
ensembles are hardly distinguishable, though the errors on the nHYP
ensemble are about 25\% lower. 

\begin{figure}
\includegraphics[scale=0.6]{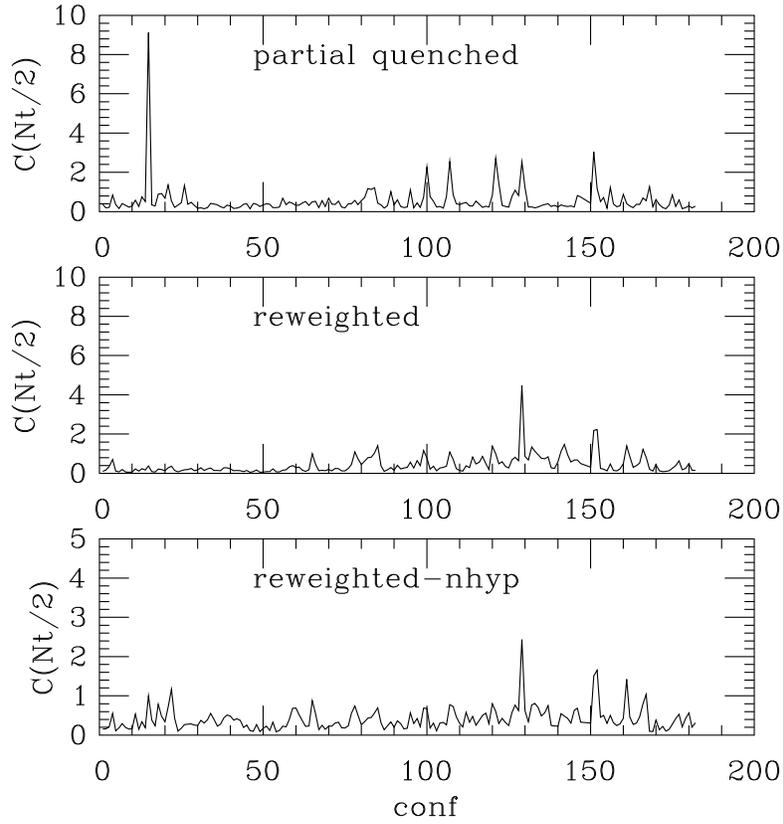}

\caption{The pseudo-scalar correlator at $t=n_{t}/2$ without reweighting and
with reweighting. The third panel shows the correlator with reweighting
that includes an nHYP plaquette term. Observe the scale difference
for the last panel. All data are for $\kappa=0.1280$.
\label{fig:The-pseudo-scalar-correlator}}

\end{figure}

In our second example we look at the pseudo-scalar correlator at $t=n_{t}/2=8$,
estimated from point-to-point propagators at a single time slice on
the individual configurations. Figure \ref{fig:The-pseudo-scalar-correlator}
shows the partial quenched and reweighted values, $C_{\pi\pi}(n_{t}/2)$
and $s_{i}C_{\pi\pi}(n_{t}/2)$, with weight factors $s_{i}$ corresponding
to a new action $S_{2}$ both with and without the nHYP plaquette
term (see Eq. (\ref{eq:nhyp_reweight})). Reweighting removes the
very large spike (corresponding to a configuration with a nearly
zero eigenmode), and reweighting with the nHYP plaquette term reduces
the fluctuations by an additional factor of two (observe the scale
difference). The reweighted data has considerably smaller statistical
fluctuations than the partially quenched one.

\section{Physical results\label{sec:Physics-results}}

Our goal in this paper is to illustrate the effectiveness of the reweighting
method. The physical results we present in this section are preliminary,
they merely serve to illustrate the power of the reweighting technique.%
\begin{table}
\begin{tabular}{|c|c|c|}
\hline 
$\kappa$  & $a\, m_{PCAC}$  & $F$(MeV)\tabularnewline
\hline
\hline 
0.1278  & 0.0119(5)  & 79(3)(4)\tabularnewline
\hline 
0.1279  & 0.0090(3)  & 79(4)(4)\tabularnewline
\hline 
0.1280  & 0.0062(3)  & 81(8)(3)\tabularnewline
\hline 
0.1281  & 0.0027(5)  & 78(10)(1)\tabularnewline
\hline
\end{tabular}

\caption{The PCAC quark mass and the low energy constant $F$ on the original
and reweighted ensembles. The first error of $F$ is statistical,
the second is systematic, due to the uncertainty of the parameter
$m\Sigma V$. \label{tab:summary}}

\end{table}

As we have mentioned in Sect.~\ref{sec:Numerical-tests}, the original
ensemble consists of 180 $16^{4}$ configurations generated with 2 degenerate
flavors of nHYP clover fermions at coupling $\beta=7.2$, $\kappa=0.1278$.
The lattice spacing is $a=0.115(3)$ as calculated from the Sommer
parameter $r_0/a=4.25(10)$ \cite{Sommer:1993ce,Hasenfratz:2001tw} using $r_{0}=0.49\,$fm,
and the PCAC quark mass is $am_{PCAC}=0.0119(5)$, translating to $m_{PCAC}\approx20\,$MeV.
From the quadratic time dependence of the axial correlator \cite{Hasenfratz:inprep},
on our volume of $V=(1.87\,$fm$)^{4}$ we estimate $m\Sigma V\approx2.1$,
in or close to the $\epsilon-$regime.

Reweighting has no observable effect on the lattice spacing, both
with standard and with nHYP reweighting $r_{0}/a=4.25$ at every $\kappa$
value, though the error increases to $0.15$ at $\kappa=0.1281$.
One reason for the relatively large error is the short time extent
of the lattices. 

Table \ref{tab:summary} lists the PCAC quark mass as obtained on
the reweighted ensembles. These values correspond to quark masses
between 5 and 20$\,$MeV. $m_{PCAC}$ shows a linear dependence on the
bare quark mass with $\kappa_{cr}=0.1282$, as shown in Figure \ref{fig:m_{PCAC}-quark}.
With reweighting we were able to decrease the quark mass by a factor
of 4. %
\begin{figure}
\includegraphics[scale=0.55]{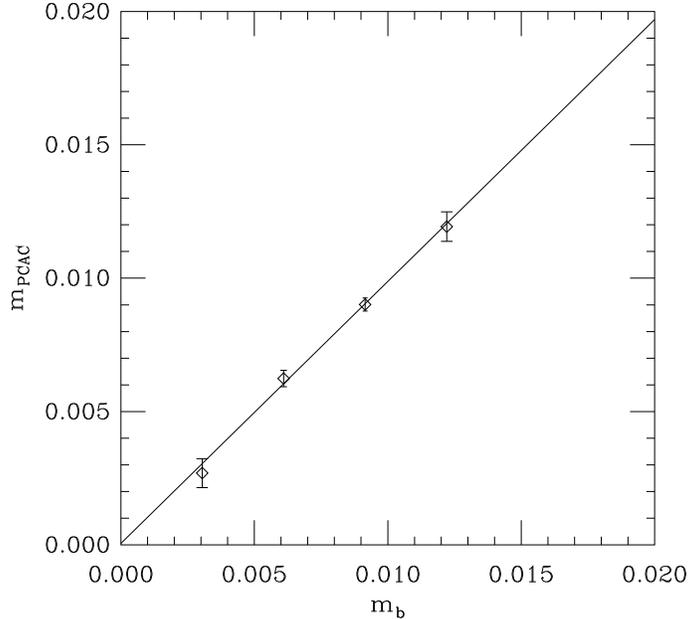}

\caption{The $m_{PCAC}$ quark mass as a function of the bare mass, $m_{q}=(1/\kappa-1/\kappa_{cr})/2$
with $\kappa_{cr}=0.1282$. \label{fig:m_{PCAC}-quark} }

\end{figure}

\begin{figure}
\includegraphics[scale=0.6]{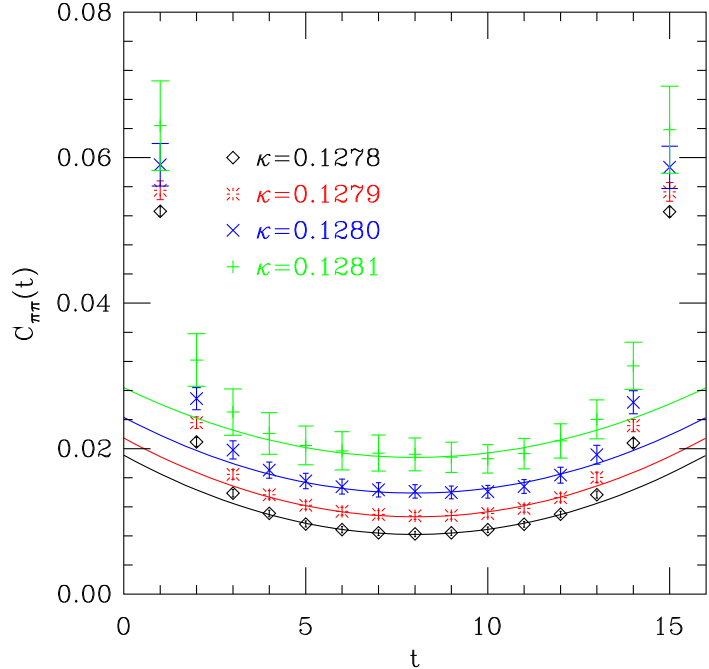}

\caption{The pseudoscalar correlator on the original and nHYP reweighted data
sets.\label{fig:The-pseudoscalar-correlator}}

\end{figure}

In the $\epsilon-$regime NLO chiral perturbation theory predicts a quadratic
form for the meson correlators. For the pseudoscalar meson the prediction
to $\mathcal{O}(\epsilon^{4})$ is \cite{Hasenfratz:1989pk,Hansen:1990un,Hansen:1990yg}
\begin{eqnarray}
G_{P}(t) & = & \frac{1}{L_{s}^{3}}\int d^{3}x\langle P(x)P(0)\rangle\nonumber \\
 & = & a_{p}+\frac{L_{t}}{L_{s}^{3}}b_{p}h_{1}\big(\frac{t}{L_{t}}\big)+
 \mathcal{O}(\epsilon^{4})\,,\label{eq:eps-pion-corr}\end{eqnarray}
 where\begin{eqnarray}
a_{p} & = & \frac{\Sigma^{2}\rho^{2}}{8}I_{1}(2\, m\Sigma V\,\rho)\label{eq:chipt_coeff}\\
b_{p} & = & \frac{\Sigma^{2}}{F^{2}}\big(1-\frac{1}{8}I_{1}(2\, m\Sigma V\,\rho)\big)\nonumber
\end{eqnarray}
 are related to the chiral low energy constants $\Sigma$ and $F$,
while \[
\rho=1+\frac{3\beta_{1}}{2F^{2}\sqrt{V}}\]
 is the shape factor with $\beta_{1}=0.14046$ for our symmetric geometry
\cite{Hasenfratz:1989pk}. $I_{1}$ can be expressed in terms of
Bessel functions, $I_{1}(u)=8Y'(u)/(uY(u))$. The function \begin{equation}
h_{1}(\tau)=\frac{1}{2}\big[(\tau-\frac{1}{2})^{2}-\frac{1}{12}\big]\label{eq:h1}\end{equation}
 describes the quadratic time dependence. Our data follows this expected
functional form in the region $t\in[4,12]$. Figure \ref{fig:The-pseudoscalar-correlator}
shows the pseudoscalar correlator on the original as well as on the
reweighted data sets with the corresponding quadratic fits. %

The fit gives the constant term $a_{p}$ with 6-8\% error, predicting
$\Sigma^{1/3}$ with 1\% accuracy, while the ratio $a_{p}/b_{p}$
has 8-30\% errors, predicting $F$ at the 4-15\% level. The large
errors are not surprising as the quadratic term measures only finite
size effects, and our lattices are not small. The errors are particularly
large at $\kappa=0.1281$ where the configuration overlap with the
original ensemble is getting small. Nevertheless we chose to present
results based on the ratio $a_{p}/b_{p}$ as it is free of renormalization
factors. To predict the low energy constant $\Sigma$ we would need
the renormalization factor of the pseudoscalar density. This calculation
is in progress and we will report the results in a forthcoming publication
\cite{Hasenfratz:inprep}. In Table $ $\ref{tab:summary} we list
the predictions for $F$ as obtained from the pseudoscalar correlator.
The first error is statistical, the second is systematic from the
uncertainty of the quantity $m\Sigma V$. The central values correspond
to $m\Sigma V=2.1$ at $\kappa=0.1278$, and at the other $\kappa$
values we rescaled this according to the PCAC quark mass. It is satisfying
that the values we obtain are consistent with each other, suggesting
that all four data sets are governed by the $\epsilon$-regime predictions.
The value is also consistent with recent $p$-regime overlap action
calculations, though somewhat smaller than overlap $\epsilon$-regime
calculations \cite{Noaki:2007es,Fukaya:2007pn,Necco:2007pr}. It
is possible to determine $F$ from the eigenvalue distribution of
the Dirac operator at imaginary chemical potential, giving consistent
results, though with large finite volume corrections \cite{DeGrand:2007tm}.
Other meson correlators can be used in similar way to predict the
low energy constants of the chiral Lagrangian.

The advantage of using Wilson-clover fermions is that it is relatively
inexpensive to create even large volume configurations. With the reweighting
technique it is possible to probe a whole range of mass values and
approach the $\epsilon-$regime without independent simulations. At
present we are running simulations on $24^{4}$ volumes at the same
lattice spacing ($L=2.8\,$fm) at approximately 8$\,$MeV quark masses.
Our tests indicate that reweighting to 2-3$\,$MeV quarks does not introduce
large statistical errors, therefore we will be able to tests the finite
volume dependence of the low energy constants.

\section{Conclusion}

Dynamical simulations with light quarks are still computationally
expensive and have to overcome technical difficulties due to large
auto-correlation time, algorithmic instability and statistical fluctuations.
In this paper we presented an alternative to direct simulations, suggesting
that reweighting in the quark mass to reach the desired light mass
value might be a better alternative. We described stochastic reweighting
and presented several improvements.

Our numerical tests show that with reweighting one can easily approach
the $\epsilon-$regime with Wilson-clover quarks and we presented
preliminary data for the low energy chiral constant $F$. Reweighting
can also provide an alternative to partial quenched studies in the
$p-$regime, since the statistical fluctuations of the reweighted
dynamical lattices can be considerably smaller than the partially
quenched data. As an example we showed how the scalar correlator,
known to become negative in partial quenched simulations, becomes
positive and follows the expected theoretical form on the reweighted
(and therefore fully dynamical) ensembles.

Reweighting might not be efficient in large volumes, or might be limited
to small mass differences. Our experience indicates that this is not
a problem on ($1.87\,$fm)$^{4}$ volumes between 20$\,$MeV and 5$\,$MeV quarks
or ($2.8\,$fm)$^{4}$ volumes between 8 and 3$\,$MeV quarks with nHYP smeared
Wilson-clover fermions.

\section{Acknowledgment}

We thank T. Degrand for useful discussions. Some of the
numerical calculations were carried out on the KAON cluster at
Fermilab and we acknowledge the allocation from USQCD/SciDac.
This research was partially supported by the US Dept. of Energy and the 
Deutsche Forschungsgemeinschaft in the SFB/TR 09.
\bibliographystyle{apsrev}
\bibliographystyle{apsrev}
\bibliography{lattice}

\end{document}